\documentclass[journal,12pt,onecolumn,draftclsnofoot,]{IEEEtran}

\usepackage{color,soul}
\usepackage{graphicx}
\usepackage{cite}
\usepackage{url}
\usepackage{amsmath, graphics,amssymb,epsfig}
\usepackage{subfigure}
\def\FigureWidth{5in}
\def\BibTeX{{\rm B\kern-.05em{\sc i\kern-.025em b}\kern-.08em
    T\kern-.1667em\lower.7ex\hbox{E}\kern-.125emX}}

\newtheorem{lemma}{Lemma}

\newcommand{\be}{\begin{equation}}

\newcommand{\ee}{\end{equation}}
\newcommand{\bea}{\begin{eqnarray}}
\newcommand{\eea}{\end{eqnarray}}
\newcommand{\bdp}{\begin{displaymath}}
\newcommand{\edp}{\end{displaymath}}
\usepackage{graphicx,amsmath}

\makeatletter
\def\hlinewd#1{%
\noalign{\ifnum0=`}\fi\hrule \@height #1 \futurelet \reserved@a\@xhline}
\newcommand{\hthickline}{\hlinewd{1.2pt}}

\makeatother

\begin{document}
\title{\huge{Minimum Throughput Maximization in UAV-Aided Wireless Powered Communication Networks}}
\author{\IEEEauthorblockN{\normalsize{Junhee Park, Hoon Lee, Subin Eom, and Inkyu Lee, \textit{Fellow}, \textit{IEEE}} \\
\IEEEauthorblockA{School of Electrical Eng., Korea University, Seoul, Korea\\
    Email: \{pjh0585, ihun1, esb777, inkyu\}@korea.ac.kr} \\\small}
}\maketitle \thispagestyle{empty}
\begin{abstract}
This paper investigates unmanned aerial vehicle (UAV)-aided wireless powered communication network (WPCN) systems where a mobile access point (AP) at the UAV serves multiple energy-constrained ground terminals (GTs). Specifically, the UAVs first charge the GTs by transmitting the wireless energy transfer (WET) signals in the downlink. Then, by utilizing the harvested wireless energy from the UAVs, the GTs send their uplink wireless information transmission (WIT) signals to the UAVs. In this paper, depending on the operations of the UAVs, we adopt two different scenarios, namely integrated UAV and separated UAV WPCNs. First, in the integrated UAV WPCN, a UAV acts as a hybrid AP in which both energy transfer and information reception are processed at a single UAV. In contrast, for the separated UAV WPCN, we consider two UAVs each of which behaves as an energy AP and an information AP independently, and thus the energy transfer and the information decoding are separately performed at two different UAVs. For both systems, we jointly optimize the trajectories of the UAVs, the uplink power control, and the time resource allocation for the WET and the WIT to maximize the minimum throughput of the GTs. Since the formulated problems are non-convex, we apply the concave-convex procedure by deriving appropriate convex bounds for non-convex constraints. As a result, we propose iterative algorithms which efficiently identify a local optimal solution for the minimum throughput maximization problems. Simulation results verify the efficiency of the proposed algorithms compared to conventional schemes.
\end{abstract}


\section{Introduction}  \label{sec:1-0Introduction}
Recently, unmanned aerial vehicles (UAVs) have been adopted in many applications such as weather monitoring and traffic control \cite{YZeng:16b}, and the usage of the UAV in wireless communication systems has drawn great attentions \cite{PZhan:11,FOno:16,YZeng:16,JLyu:17,MAlzenad:17,MAzari:17,YZeng:17,QWu:17,MChen:17,SJeong:17}.
Compared to conventional networks where APs are fixed at given locations, wireless communication networks employing a UAV-mounted access point (AP) exhibit cost-efficiency and deployment flexibility.
Moreover, the mobility of the UAV can provide an opportunity for the wireless networks to enhance the system capacity.

In \cite{PZhan:11,FOno:16,YZeng:16}, UAV-enabled relaying channels were studied where UAVs act as mobile relays which forward the information of sources to destinations located on the ground. For the UAV relay networks, deployment and direction control problems were investigated in \cite{PZhan:11}, and the work in \cite{FOno:16} minimized the network outage probability when the UAV trajectory is given as a circular path.
The authors in \cite{YZeng:16} solved the throughput maximization problem by optimizing the source and the relay transmit power allocation along with the UAV relay trajectory.
In addition, UAVs have been employed as mobile base stations in various wireless networks \cite{JLyu:17,MAlzenad:17,MAzari:17,YZeng:17,QWu:17}.
The mobile base station placement problems were investigated in \cite{JLyu:17} and \cite{MAlzenad:17} in order to maximize the overall wireless coverage. In \cite{MAzari:17}, analytical expressions for the optimal UAV height were derived to minimize the outage probability of air-to-ground links.
The authors in \cite{YZeng:17} focused on the theoretical energy consumption modeling for UAVs, and proposed trajectory optimization methods for maximizing the energy efficiency of a UAV.
Also, the trajectories of multiple UAVs were examined in \cite{QWu:17} to maximize the minimum throughput performance of multiple ground terminals (GTs). Moreover, UAV-aided caching and mobile cloud computing systems were researched in \cite{MChen:17} and \cite{SJeong:17}, respectively.

In the meantime, energy harvesting (EH) techniques based on radio frequency (RF) signals have been considered as promising solutions for extending the lifetime of battery-limited wireless devices \cite{RZhang:13,QShi:14,HLee:15,CSong:16,JKim:17,HJu:14,LLiu:14,HKim:16,HLee:16}. By utilizing wireless energy transfer (WET) and wireless information transmission (WIT), the RF-based EH methods have been studied for traditional wireless communications, and wireless powered communication networks (WPCN) protocols have been widely investigated in recent literature \cite{HJu:14,LLiu:14,HKim:16,HLee:16}.

Particularly, in the WPCN, a hybrid access point (H-AP) sends wireless energy via the RF signals to energy-constrained devices in the downlink WET phase. In the subsequent uplink WIT phase, the devices transmit their information signals to the H-AP by using the harvested energy.
In \cite{HJu:14}, throughput maximization problems were introduced for the WPCN by optimizing the time resource allocated to users under the harvest-then-transmit protocol.
The authors in \cite{LLiu:14} proposed the multi-antenna energy beamforming and time allocation algorithms to maximize the minimum throughput performance.
The sum-rate maximization problems with a full-duplex H-AP were investigated in \cite{HKim:16} for orthogonal frequency division multiplexing, and the precoding methods for the multiple-input multiple-output WPCN was provided in \cite{HLee:16}.
Note that these works were restricted to a static H-AP setup, and thus it would suffer from the `doubly near-far' problem \cite{HJu:14}, which is induced by the doubly distance-dependent signal attenuation both in the downlink and the uplink.

Recently, there have been several works combining mobile vehicle techniques and the WPCN \cite{YShi:11,LXie:12,SGuo:14,MZhao:14,TLi:16,LFu:16,JXu:17}.
For the magnetic resonant based WET, \cite{YShi:11,LXie:12,SGuo:14,MZhao:14} considered wireless charging vehicles which travel the networks to supply power to wireless sensors.
However, due to short charging coverage of the magnetic resonance technique, the vehicles should stay quite a while to transfer energy to nearby sensors.
To overcome this limitation, the authors in \cite{TLi:16} adopted the RF-based WET methods to UAV-aided WPCN where a UAV flies towards a GT to transmit the RF energy signal and receive uplink data. However, only a single GT case was considered in \cite{TLi:16} under a fixed line trajectory setup without optimizing the traveling path of the UAV.
The works in \cite{LFu:16} and \cite{JXu:17} also examined the UAV-enabled WET systems, but they did not take into account the communications of GTs.

In this paper, we investigate the UAV-aided WPCN where multiple energy-constrained GTs are served by UAVs with arbitrary trajectories. Depending on the roles of the UAVs, we classify the UAV WPCN into two categories: integrated UAV and separated UAV WPCNs. First, in the integrated UAV WPCN, a single UAV behaves as an H-AP which broadcasts the RF energy signal to the GTs in the downlink WET phase and decodes the information from the GTs in the uplink WIT phase. In contrast, in the separated UAV WPCN, the WET and WIT operations are assigned to two different UAVs separately. In both systems, we adopt a time division multiple access (TDMA) based harvest-then-transmit protocol in \cite{HJu:14} where the WET of the UAVs and the WIT at the GTs are performed over orthogonal time resources.

In our proposed systems, we jointly optimize the trajectories of the UAVs, the uplink power control at the GTs, and the time resource allocation with the aim of maximizing the minimum throughput performance among the GTs. Since the location of the UAVs changes continuously, the time resource allocation in the UAV WPCN is totally different from that of the conventional WPCN with static H-APs. Also, compared to \cite{TLi:16} where the trajectory of the UAV is restricted to a straight line, our systems consider a general traveling path optimization problem without any constraints on the UAV trajectory.

As the problem is non-convex, we propose iterative algorithms to obtain the local optimal solution by applying the alternating optimization method.
To be specific, we first jointly optimize the trajectory of the UAVs and the uplink power of the GTs with given time allocation, and then update the time resource allocation solution by fixing other variables. First, to find the trajectory and the uplink power control solution, the concave-convex procedure (CCCP) framework \cite{LAn:05}\cite{YSun:17} is employed which successively solves approximated convex problems of the original problem. Next, the time allocation solution can be determined by applying linear programming (LP). The convergence and the local optimality of the proposed algorithms are then mathematically proved. From numerical results, we demonstrate that the proposed algorithms substantially improve the performance of the UAV WPCN compared to conventional schemes.

The rest of this paper is organized as follows: Section \ref{sec:2-0system model} explains a system model for the UAV WPCN and formulates the minimum throughput maximization problems. In Sections \ref{sec:3-0PropSol_IS} and \ref{sec:4-0PropSol_SS}, we propose efficient algorithms for the integrated UAV and the separated UAV systems, respectively. Section \ref{sec:5-0simulation} presents simulation results for the proposed algorithms and compares the performance with conventional schemes. Finally, the paper is terminated in Section \ref{sec:6-0conclusion} with conclusions.

Throughout this paper, normal and boldface letters represent scalar quantities and column vectors, respectively. We denote the Euclidean space of dimension $n$ as $\mathbb{R}^{n}$, and $(\cdot)^{T}$ indicates the transpose operation. Also, $|\cdot|$ and $\|\cdot\|$ stand for the absolute value and the 2-norm, respectively.

\section{System Model} \label{sec:2-0system model}

\begin{figure}
\centering
    \subfigure[Integrated UAV WPCN]{
        \includegraphics[width=3.1in]{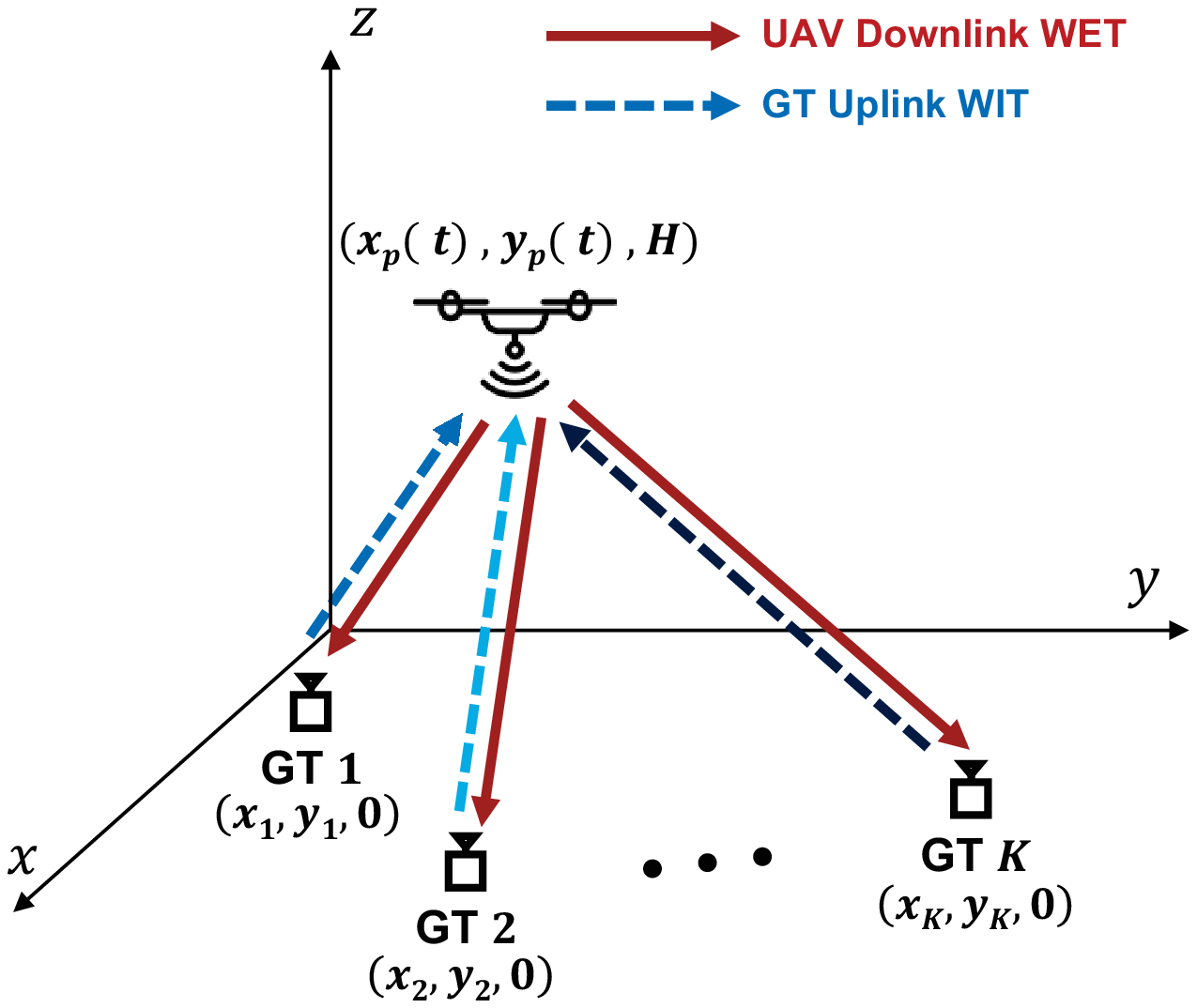}
    \label{figure:IS}}
    \subfigure[Separated UAV WPCN]{
        \includegraphics[width=3.1in]{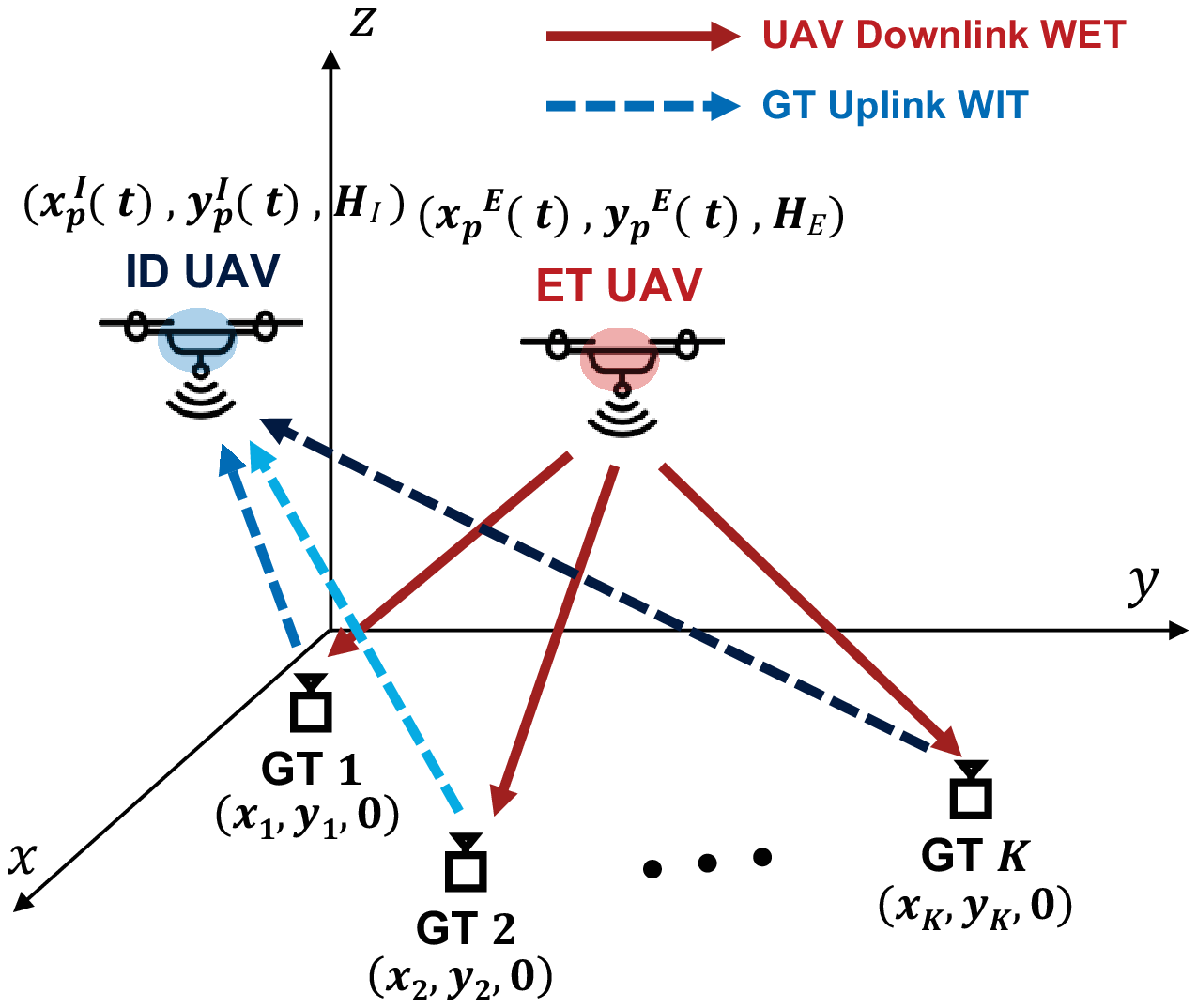}
    \label{figure:SS}}
    \caption{Schematic diagrams of UAV-aided WPCNs.}
    \label{figure:system_model}
\end{figure}

As shown in Figure \ref{figure:system_model}, we consider a UAV-aided WPCN where $K$ single antenna GTs are supported by single antenna UAVs which transmit and receive the RF signals. It is assumed that the GTs do not have any embedded power supplies, while the UAVs are equipped with stable and constant power sources. To communicate with the GTs, the UAVs travel through the area of interest while transferring energy to the GTs in the downlink. By utilizing the harvested energy from the UAV, the GTs send their information in the uplink. We assume that the UAVs fly at a constant altitude of $H$ with the maximum speed $v_{\text{max}}$ for the time period $T$, whereas all the GTs are fixed at given locations.

Depending on the operations of the UAVs, we classify the UAV-aided WPCN into two categories. First, in the integrated UAV WPCN illustrated in Figure \ref{figure:IS}, a single UAV transmits energy and collects data of the GTs. Thus, the UAV in the integrated UAV WPCN acts as an H-AP in the conventional WPCN \cite{HJu:14}. Second, in the separated UAV WPCN in Figure \ref{figure:SS}, the WET and the WIT are independently performed at two different UAVs. Therefore, each UAV in the separated system is dedicated to the energy transferring (ET) or the information decoding (ID). In the following, we present the system model for both UAV WPCN systems.

\subsection{Integrated UAV WPCN} \label{sec:2-1_IS}
Let us denote $\mathbf{p}(t)=[x_{p}(t),y_{p}(t)]^T$ as the position of the UAV at time instant $t\in[0,T]$ and $\mathbf{u}_{k}=[x_{k},y_{k}]^T$ as the location of GT $k \in \mathcal{K} \triangleq \{1,...,K\}$, which is assumed to be known to the UAV in advance. For ease of analysis, the total time period $T$ is equally divided into $N$ time slots as in \cite{QWu:17}, where the number of time slots $N$ is chosen as a sufficiently large number such that the distance between the UAV and the GTs within each time slot can be considered approximately static.

Therefore, the trajectory of the UAV can be represented by a sequence of locations $\{\mathbf{p}[n]\}$ at each time slot $n \in \mathcal{N} \triangleq \{1,...,N\}$ as
\bea
    \mathbf{p}[n] \triangleq \mathbf{p}(n\delta_{N})=[x_{p}(n\delta_{N}),y_{p}(n\delta_{N})]^T, \nonumber
\eea
where $\delta_{N} \triangleq T/N$ indicates the length of the time slot.
Since we consider the discrete time trajectory $\mathbf{p}[n]\ \mbox{for} \ n \in \mathcal{N}$, the maximum speed constraint can be expressed as
\bea
    \|\mathbf{p}[n]-\mathbf{p}[n-1]\| \leq \delta_{N}v_{\text{max}},\ \mbox{for} \ \ n\in\mathcal{\hat{N}}\triangleq\{2,...,N\}. \nonumber\label{eq:SpeedConstraint}
\eea

For the air-to-ground channel between the UAV and the GTs, the deterministic propagation model is adopted in this paper which assumes the line-of-sight links without the Doppler effect \cite{YZeng:16},\cite{YZeng:17},\cite{SJeong:17}. Then, the average channel power gain $\gamma_{k}[n]$ between the UAV and GT $k$ at time slot $n$ is given by
\bea
    \gamma_{k}[n]= \frac{g_{0}}{\|\mathbf{p}[n]-\mathbf{u}_{k}\|^2 + H^2},\ \ \mbox{for} \ \ n\in\mathcal{N} \ \mbox{and} \  k\in\mathcal{K}, \nonumber\label{eq:ChannelGain}
\eea
where $g_{0}$ denotes the reference channel gain at distance of 1 meter.

\begin{figure}
\begin{center}
\includegraphics[width=\FigureWidth]{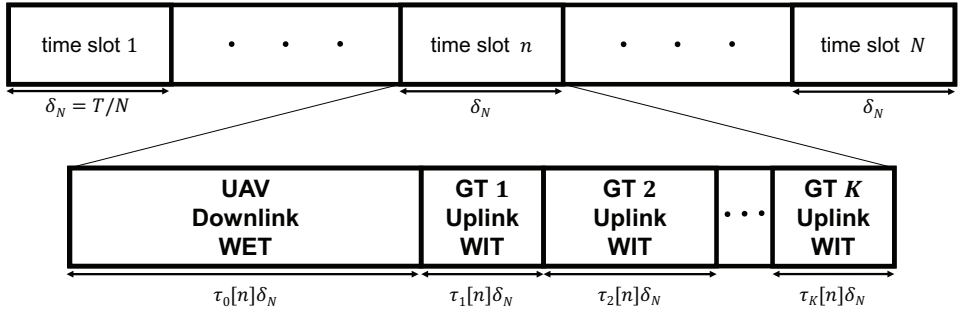}
\end{center}
\vspace{-5mm}
\caption{Protocol structure for UAV-aided WPCN.}
\label{figure:system protocol}
\end{figure}

Next, we explain the transmission protocol for the UAV-aided WPCN. As shown in Figure \ref{figure:system protocol}, we divide each time slot $n$ into $K+1$ subslots, where the 0-th subslot of duration $\tau_{0}[n]\delta_{N}$ is allocated to the dedicated downlink WET and the $k$-th subslot of duration $\tau_{k}[n]\delta_{N}$ for $k\in\mathcal{K}$ is assigned to the uplink WIT of GT $k$. Note that the variable $\tau_{k}[n]$ accounts for the time durations at the $k$-th subslot in time slot $n$. Thus we have the following constraints on the time resource allocation variable $\tau_{k}[n]$ as
\bea
    &&0\leq \tau_{k}[n] \leq 1,\ \ \mbox{for} \ \ n\in\mathcal{N} \ \mbox{and} \ k\in\mathcal{\breve{K}},  \nonumber\\
    &&\sum_{k=0}^{K}{\tau_{k}[n]} \leq 1,\ \ \mbox{for} \ \ n\in\mathcal{N}, \nonumber
\eea
where $\mathcal{\breve{K}} \triangleq \mathcal{K}\cup\{0\}$.

Now, we describe the WET and the WIT procedures of the integrated UAV WPCN. At the 0-th subslot of each time slot, the UAV broadcasts the wireless energy signals with the transmit power $P^{\text{DL}}$. Then, the harvested energy $E_{k}[n]$ of GT $k$ at time slot $n$ can be written as
\bea
    E_{k}[n] &\triangleq& \tau_{0}[n]\delta_{N}\cdot\zeta_k\gamma_{k}[n]P^{\text{DL}} \nonumber \\
            &=& \tau_{0}[n]\delta_{N}\frac{\zeta_{k}g_{0}P^{\text{DL}}}{\|\mathbf{p}[n]-\mathbf{u}_{k}\|^2 + H^2},\ \ \mbox{for} \ \ n\in\mathcal{N} \ \mbox{and} \ k\in\mathcal{K},  \nonumber \label{eq:HarvestedEnergyPerUnit}
\eea
where $\zeta_k \in (0,1]$ stands for the energy harvesting efficiency of GT $k$. For simplicity, we assume that all the GTs have the same energy harvesting efficiency, i.e., $\zeta_{k}=\zeta$ for $k\in\mathcal{K}$.

Due to the processing delay of EH circuits at the GTs, the harvested energy $E_{k}[n]$ may not be available at time slot $n$. Hence, GT $k$ only can utilize $E_{k}[n]$ at the future time slots $n+1,n+2,...,N$.
Defining $P_{k}^{\text{UL}}[n]$ as the uplink transmit power of GT $k$ at time slot $n$, the available energy $\tilde{E}_{k}[n]$ at time slot $n$ of GT $k$ can be expressed as
\bea
    \tilde{E}_{k}[n]=\sum_{i=1}^{n-1}{E_{k}[i]} - \sum_{i=1}^{n-1}{\tau_{k}[i]\delta_{N}P_{k}^{\text{UL}}[i]},  \nonumber\label{eq:AvailableEnergyNotation}
\eea
where the first and the second terms represent the cumulative harvested energy and the consumed energy of GT $k$ during the past time slots for $i=1,2,...,n-1$, respectively. As a result, the uplink power constraint for GT $k$ at time slot $n$ is given as
\bea
    \tau_{k}[n]\delta_{N}P_{k}^{\text{UL}}[n] \leq \tilde{E}_{k}[n],\ \ \mbox{for} \ \ n\in\mathcal{\hat{N}} \ \mbox{and} \ k\in\mathcal{K},\label{eq:AvailableEnergy}
\eea
where we have $P_{k}^{\text{UL}}[1]=0$ due to the EH circuit delay.

Also, the instantaneous throughput $R_{k}[n]$ of GT $k$ at time slot $n$ can be obtained as
\bea
    R_{k}[n] &\triangleq& \log_{2}\bigg(1+\frac{\eta_{k}\gamma_{k}[n]P_{k}^{\text{UL}}[n]}{\sigma^2}\bigg) \nonumber \\
                &=& \log_{2}\bigg(1+ \frac{g_{0}\eta_{k}}{\sigma^2}\frac{P_{k}^{\text{UL}}[n]}{\|\mathbf{p}[n]-\mathbf{u}_{k}\|^2 + H^2}\bigg),\ \ \mbox{for} \ \ n\in\mathcal{N} \ \mbox{and} \ k\in\mathcal{K}, \nonumber\label{eq:ThroughputPerUnit}
\eea
where $\eta_{k} \in (0,1]$ is a portion of the stored energy used for the uplink information transmission at GT $k$. For simplicity, we assume $\eta_{k}=\eta$ for $k\in\mathcal{K}$. Then, the average throughput $R_{k}$ of GT $k$ for the time period $T$ can be written by
\bea
    R_{k} &\triangleq& \frac{1}{T}\delta_{N}\sum_{n=2}^{N}\tau_{k}[n]R_{k}[n] \nonumber\\
            &=&\frac{1}{N}\sum_{n=2}^{N}\tau_{k}[n]\log_{2}\bigg(1+\frac{g_{0}\eta}{\sigma^2}\frac{P_{k}^{\text{UL}}[n]}{\|\mathbf{p}[n]-\mathbf{u}_{k}\|^2 + H^2}\bigg),\ \ \mbox{for} \ \ k\in\mathcal{K}.
\eea

In this paper, we aim to maximize the minimum average throughput of the GTs by jointly optimizing the UAV trajectory $\{\mathbf{p}[n]\}$, the uplink power control $\{P_{k}^{\text{UL}}[n]\}$ at the GTs, and the time resource allocation variables $\{\tau_{k}[n]\}$. Denoting $R_{\text{min}}$ as the minimum throughput of the GTs, the optimization problem can be formulated as
\bea
    \lefteqn{\mbox{(P1) }\max_{R_{\text{min}}, \{P_{k}^{\text{UL}}[n]\},\{\mathbf{p}[n]\},\{\tau_{k}[n]\}}R_{\text{min}}} \label{eq:P1} \nonumber \\
    \lefteqn{s.t. } &&
    \frac{1}{N}\sum_{n=2}^{N}\tau_{k}[n]\log_{2}\bigg(1+\frac{g_{0}\eta}{\sigma^2}\frac{P_{k}^{\text{UL}}[n]}{\|\mathbf{p}[n]-\mathbf{u}_{k}\|^2 + H^2}\bigg)\geq R_{\text{min}},\ \ \mbox{for} \ k\in\mathcal{K},\label{eq:P1-Throughput}\\ 
    &&\sum_{i=2}^{n}\tau_{k}[i]P_{k}^{\text{UL}}[i]
                        \leq \sum_{i=1}^{n-1}\frac{\tau_{0}[i]g_{0}\zeta_{k}P^{\text{DL}}}{\|\mathbf{p}[i]-\mathbf{u}_{k}\|^2 + H^2},\ \ \mbox{for} \ \ n\in\mathcal{\hat{N}} \ \mbox{and} \ k\in\mathcal{K},\label{eq:P1-AvailEn}  \\ 
    &&\|\mathbf{p}[n]-\mathbf{p}[n-1]\| \leq \delta_{N} v_{\text{max}},\ \ \mbox{for} \ \ n\in\mathcal{\hat{N}},\label{eq:P1-Speed}  \\ 
    &&\|\mathbf{p}[N]-\mathbf{p}[1]\| \leq \delta_{N} v_{\text{max}},\label{eq:P1-InitFinalLoc}  \\  
    &&0 \leq P_{k}^{\text{UL}}[n] \leq P_{\text{max}}^{\text{UL}},\ \ \mbox{for} \ \ n\in\mathcal{N} \ \mbox{and} \ k\in\mathcal{K},\label{eq:P1-MaxULPower}     \\
    &&0 \leq \tau_{k}[n] \leq 1,\ \ \mbox{for} \ \ n\in\mathcal{N} \ \mbox{and} \ k\in\mathcal{\breve{K}}, \label{eq:P1-tau1} \\
    &&\sum_{k=0}^{K}\tau_{k}[n]=1\ \ \mbox{for} \ \ n\in\mathcal{N}, \label{eq:P1-tau2}
\eea
where the uplink energy constraint in (\ref{eq:P1-AvailEn}) is derived from (\ref{eq:AvailableEnergy}), (\ref{eq:P1-InitFinalLoc}) indicates the periodical constraint that the UAV needs to get back to the starting position after one time period $T$ \cite{QWu:17}\footnote{Depending on the application, one may want to determine the initial location and the final location of the UAV in advance. In this case, we can simply add constraints on $\mathbf{p}[0]$ and $\mathbf{p}[N]$ and discard the constraint in (\ref{eq:P1-InitFinalLoc}).}, and (\ref{eq:P1-MaxULPower}) is the peak uplink power constraint. One can check that (P1) is non-convex due to the constraints in (\ref{eq:P1-Throughput}) and (\ref{eq:P1-AvailEn}), and therefore it is not straightforward to obtain the globally optimal solution.

\subsection{Separated UAV WPCN} \label{sec:2-2_SS}
In the separated UAV WPCN, we design the trajectories of two different UAVs, i.e., ID UAV and ET UAV. Let us define $\mathbf{p}_{\scriptscriptstyle{\text{I}}}[n] \in \mathbb{R}^2$ and $\mathbf{p}_{\scriptscriptstyle{\text{E}}}[n] \in \mathbb{R}^2$ as the position of the ET UAV and the ID UAV at time slot $n$, respectively.
Similar to the integrated UAV WPCN, we adopt the TDMA protocol in Figure \ref{figure:system protocol}.
Then, the uplink energy constraint of GT $k$ at time slot $n$ and the average throughput of GT $k$ $R_{k,\text{S}}$ can be respectively expressed as
\bea
    \tau_{k}[n]\delta_{N} P_{k}^{\text{UL}}[n] &\leq& \sum_{i=1}^{n-1}{\Bigg( \tau_{0}[n]\delta_{N}\frac{\zeta_{k}g_{0}P^{\text{DL}}}{\|\mathbf{p}_{\scriptscriptstyle{\text{E}}}[i]-\mathbf{u}_{k}\|^2 + H_{\scriptscriptstyle{\text{E}}}^2} - \tau_{k}[i]\delta_{N} P_{k}^{\text{UL}}[i]\Bigg)},\nonumber\label{eq:AvailableEnergy_SS} \\
    R_{k,\text{S}} &\triangleq& \frac{1}{N}\sum_{n=2}^{N}\tau_{k}[n]\log_{2}\bigg(\frac{g_{0}\eta}{\sigma^2}\frac{P_{k}^{\text{UL}}[n]}{\|\mathbf{p}_{\scriptscriptstyle{\text{I}}}[n]-\mathbf{u}_{k}\|^2 + H_{\scriptscriptstyle{\text{I}}}^2}\bigg), \nonumber
\eea
where $H_{\scriptscriptstyle{\text{I}}}$ and $H_{\scriptscriptstyle{\text{E}}}$ stand for the flight altitude of the ID UAV and the ET UAV, respectively.

Thus, the minimum throughput maximization problem for the separated UAV WPCN is given as
\bea
    \lefteqn{\mbox{(P2) }\max_{\substack{R_{\text{min}}, \{P_{k}^{\text{UL}}[n]\},\{\mathbf{p}_{\scriptscriptstyle{\text{I}}}[n]\}, \label{eq:P2}\\ \{\mathbf{p}_{\scriptscriptstyle{\text{E}}}[n]\},\{\tau_{k}[n]\}}}R_{\text{min}}} \nonumber\\
    \lefteqn{s.t. } &&
    \frac{1}{N}\sum_{n=2}^{N}\tau_{k}[n]\log_{2}\bigg(1+\frac{g_{0}\eta}{\sigma^2}\frac{P_{k}^{\text{UL}}[n]}{\|\mathbf{p}_{\scriptscriptstyle{\text{I}}}[n]-\mathbf{u}_{k}\|^2 + H_{\scriptscriptstyle{\text{I}}}^2}\bigg) \geq R_{\text{min}},\ \ \mbox{for} \ \ k\in\mathcal{K},\label{eq:P2-Throughput}\\ 
    &&\sum_{i=2}^{n}\tau_{k}[i]P_{k}^{\text{UL}}[i]
                        \leq \sum_{i=1}^{n-1}\frac{\tau_{0}[i]g_{0}\zeta_{k}P^{\text{DL}}}{\|\mathbf{p}_{\scriptscriptstyle{\text{E}}}[i]-\mathbf{u}_{k}\|^2 + H_{\scriptscriptstyle{\text{E}}}^2},\ \mbox{for} \ \ n\in\mathcal{\hat{N}} \ \mbox{and} \ k\in\mathcal{K},\label{eq:P2-AvailEn}  \\ 
    &&\|\mathbf{p}_{\scriptscriptstyle{\scriptscriptstyle{x}}}[n]-\mathbf{p}_{\scriptscriptstyle{\scriptscriptstyle{x}}}[n-1]\| \leq \delta_{N} v_{\text{max}}^{\scriptscriptstyle{x}},\ \mbox{for} \ x\in\{\text{I},\text{E}\} \ \mbox{and} \ n\in\mathcal{\hat{N}},\label{eq:P2-SpeedIAP}  \\
    &&\|\mathbf{p}_{\scriptscriptstyle{x}}[N] - \mathbf{p}_{\scriptscriptstyle{x}}[1]\| \leq \delta_{N} v_{\text{max}}^{\scriptscriptstyle{x}}, \ \mbox{for} \ x\in\{\text{I},\text{E}\},\label{eq:P2-InitFinalLoc}  \\
    &&0 \leq P_{k}^{\text{UL}}[n] \leq P_{\text{max}}^{\text{UL}},\ \ \mbox{for} \ \ n\in\mathcal{N} \ \mbox{and} \ k\in\mathcal{K},\label{eq:P2-MaxULPower}     \\
    && \mbox{(\ref{eq:P1-tau1}) - (\ref{eq:P1-tau2})}, \nonumber
\eea
where $v_{\text{max}}^{\text{I}}$ and $v_{\text{max}}^{\text{E}}$ represent the maximum speed of the ID UAV and the ET UAV, respectively.
This problem is also non-convex due to the constraints (\ref{eq:P2-Throughput}) and (\ref{eq:P2-AvailEn}). In the following sections, we present efficient approaches for solving (P1) and (P2).

\section{Proposed Solution for Integrated UAV WPCN} \label{sec:3-0PropSol_IS}
In this section, we propose an iterative algorithm for (P1) which yields a local optimal solution. To this end, we employ the alternating optimization framework which first finds a solution for the trajectory $\{\mathbf{p}[n]\}$ and the uplink power $\{P_{k}^{\text{UL}}[n]\}$ with given time resource allocation $\{\tau_{k}[n]\}$, and then computes $\{\tau_{k}[n]\}$ by fixing $\{\mathbf{p}[n]\}$ and $\{P_{k}^{\text{UL}}[n]\}$.

\subsection{Joint Trajectory and Uplink Power Optimization} \label{sec:3-1_subprob1}
For a given time resource allocation $\{\tau_{k}[n]\}$, (P1) can be simplified as
\bea
    \lefteqn{\max_{R_{\text{min}}, \{P_{k}^{\text{UL}}[n]\},\{\mathbf{p}[n]\}}R_{\text{min}}\quad \quad \quad \quad \quad \quad \quad \quad \quad \quad \quad \quad \quad } \label{P1_sub1} \\
    &s.t. \quad  \quad \mbox{(\ref{eq:P1-Throughput}) - (\ref{eq:P1-MaxULPower})}. \nonumber
\eea
Problem (\ref{P1_sub1}) is still non-convex due to the constraints in (\ref{eq:P1-Throughput}) and (\ref{eq:P1-AvailEn}). To tackle this difficulty, let us first introduce auxiliary variables $\{z_{k}[n]\}$ such that $\|\mathbf{p}[n]-\mathbf{u}_{k}\|^2 \leq z_{k}[n]$ for $k\in\mathcal{K}$ and $n\in\mathcal{N}$. Then, the left hand side (LHS) of (\ref{eq:P1-Throughput}) and the right hand side (RHS) of (\ref{eq:P1-AvailEn}) are respectively lower-bounded by
\bea
    \frac{1}{N}\sum_{n=2}^{N}\tau_{k}[n]\log_{2}\bigg(1+\frac{\frac{g_{0}\eta}{\sigma^2}P_{k}^{\text{UL}}[n]}{\|\mathbf{p}[n]-\mathbf{u}_{k}\|^2 + H^2}\bigg) &\geq& \frac{1}{N}\sum_{n=2}^{N}\tau_{k}[n]\log_{2}\bigg(1+\frac{\frac{g_{0}\eta}{\sigma^2}P_{k}^{\text{UL}}[n]}{z_{k}[n] + H^2}\bigg), \nonumber\label{eq:ThroughputLB1} \\
    \sum_{i=1}^{n-1}\tau_{0}[i]\frac{g_{0}\zeta P^{\text{DL}}}{\|\mathbf{p}[i]-\mathbf{u}_{k}\|^2 + H^2} &\geq& \sum_{i=1}^{n-1}\tau_{0}[i]\frac{g_{0}\zeta P^{\text{DL}}}{z_{k}[i] + H^2}. \nonumber \label{eq:HarvEnLB1}
\eea
For these bounds, we can construct an equivalent problem for (\ref{P1_sub1}) based on the following lemma.

\begin{lemma} \label{proposition:equiv}
The optimal solution for the problem (\ref{P1_sub1}) can be obtained by solving the following optimization problem:
\bea
    \lefteqn{\mbox{(P1.1) }\max_{R_{\text{min}}, \{P_{k}^{\text{UL}}[n]\},\{\mathbf{p}[n]\},\{z_{k}[n]\}}R_{\text{min}}}  \nonumber \\
    \lefteqn{s.t. }
    &&\frac{1}{N}\sum_{n=2}^{N}\tau_{k}[n]\log_{2}\bigg(1+\frac{g_{0}\eta}{\sigma^2}\frac{P_{k}^{\text{UL}}[n]}{z_{k}[n] + H^2}\bigg)\geq R_{\text{min}},\ \mbox{for} \ \ k\in\mathcal{K}, \label{eq:P1eq-Throughput}\\
    &&\sum_{i=2}^{n}\tau_{k}[i]P_{k}^{\text{UL}}[i] \leq \sum_{i=1}^{n-1}\tau_{0}[i]\frac{g_{0}\zeta_{k}P^{\text{DL}}}{z_{k}[i] + H^2},\ \mbox{for} \ \ n\in\mathcal{\hat{N}} \ \mbox{and} \ k\in\mathcal{K}, \label{eq:P1eq-AvailEn}   \\
    &&\|\mathbf{p}[n]-\mathbf{u}_{k}\|^2 \leq z_{k}[n],\ \mbox{for} \ \ n\in\mathcal{N} \ \mbox{and} \ k\in\mathcal{K}, \label{eq:P1eq-Aux} \\
    && \mbox{(\ref{eq:P1-Speed}) - (\ref{eq:P1-MaxULPower})}. \nonumber
\eea
\end{lemma}
\begin{IEEEproof}
First, let $R_{\text{min}}^{\ast}$ and $\tilde{R}_{\text{min}}$ denote the optimal value of problem (\ref{P1_sub1}) and (P1.1), respectively. Then it can easily be checked that $R_{\text{min}}^{\ast} \geq \tilde{R}_{\text{min}}$, where the equality holds when $z_{k}[n]=\|\mathbf{p}[n]-\mathbf{u}_{k}\|^2$, $\forall n$ and $\forall k$. Next, by contradiction, we will prove that the optimum of (P1.1) can be attained when $z_{k}[n]=\|\mathbf{p}[n]-\mathbf{u}_{k}\|^2$. Suppose that there exists at least one $z_{k}[n]$ satisfying $z_{k}[n] > \|\mathbf{p}[n]-\mathbf{u}_{k}\|^2$ at the optimum of (P1.1) and denote a set of such $k$ as $\mathcal{K}'\subset\mathcal{K}$. If the equality holds in (\ref{eq:P1eq-Throughput}) for $k'\in\mathcal{K}'$, the minimum throughput $\tilde{R}_{\text{min}}$ can be increased by reducing $z_{k'}[n]$ so that constraints (\ref{eq:P1eq-Throughput}) and (\ref{eq:P1eq-AvailEn}) hold with equality. This contradicts the assumption. Even if the equality does not hold in (\ref{eq:P1eq-Throughput}) for $k'$ at the optimum, decreasing $z_{k'}[n]$ does not affect the minimum throughput $\tilde{R}_{\text{min}}$. Therefore, for all these cases, we can always find the optimal $z_{k}[n]$ for (P1.1) satisfying $z_{k}[n]=\|\mathbf{p}[n]-\mathbf{u}_{k}\|^2$.
As a result, the optimal solution of (\ref{P1_sub1}) can be equivalently obtained by solving (P1.1).
\end{IEEEproof}

Still, (P1.1) is non-convex in general. Thus, we provide the CCCP \cite{LAn:05} approach to address (P1.1). First, we consider the throughput constraint in (\ref{eq:P1eq-Throughput}). By using a first-order Taylor approximation at $z_{k}[n]=\hat{z}_{k}[n]$, we can derive a concave lower bound for the LHS of (\ref{eq:P1eq-Throughput}) as
\bea
            \log_{2}\bigg(1+\frac{g_{0}\eta}{\sigma^2}\frac{P_{k}^{\text{UL}}[n]}{z_{k}[n] + H^2}\bigg)
            &\geq& \log_{2}\bigg(\frac{z_{k}[n] + H^2+\frac{g_{0}\eta}{\sigma^2}P_{k}^{\text{UL}}[n]}{\hat{z}_{k}[n] + H^2}\bigg)-\frac{z_{k}[n]+H^2}{\hat{z}_{k}[n] + H^2}+1 \nonumber   \\
            &\triangleq& R_{k}^{L}[n](z_{k}[n],P_{k}^{\text{UL}}[n]\mid \hat{z}_{k}[n]). \label{Rlower}
\eea

Note that $R_{k}^{L}[n](z_{k}[n],P_{k}^{\text{UL}}[n]\mid \hat{z}_{k}[n])$ is a jointly concave function with respect to $z_{k}[n]$ and $P_{k}^{\text{UL}}[n]$, and gives a tight lower bound in which equality holds at $\hat{z}_{k}[n]=z_{k}[n]$.
In a similar way, the RHS of constraint (\ref{eq:P1eq-AvailEn}), which is convex with respect to $z_{k}[n]$, can be lower-bounded by
\bea
    \tau_{0}[n]\delta_{N}\frac{\zeta_{k}g_{0}P^{\text{DL}}}{z_{k}[n] + H^2}
            &\geq& \frac{\tau_{0}[n]\delta_{N}\zeta g_{0}P^{\text{DL}}}{\hat{z}_{k}[n] + H^2}
            \bigg(   2 - \frac{z_{k}[n]+H^2}{\hat{z}_{k}[n] + H^2}   \bigg) \nonumber \\
            &\triangleq& E_{k}^{L}[n](z_{k}[n] \mid \hat{z}_{k}[n]). \label{Elower}
\eea

With (\ref{Rlower}) and (\ref{Elower}) at hand, an approximated convex problem for (P1.1) with given $\hat{z}_{k}[n]$ can be formulated as
\bea
    \lefteqn{\mbox{(P1.1A) }\max_{R_{\text{min}}, \{P_{k}^{\text{UL}}[n]\},\{\mathbf{p}[n]\},\{z_{k}[n]\}}R_{\text{min}}} \label{eq:P1-1A} \nonumber \\
    \lefteqn{s.t. }
    &&\frac{1}{N}\sum_{n=2}^{N}\tau_{k}[n]R_{k}^{L}[n](z_{k}[n],P_{k}^{\text{UL}}[n]\mid \hat{z}_{k}[n]) \geq R_{\text{min}},\ \mbox{for} \ \ k\in\mathcal{K}, \label{eq:P1approx-Throughput}\\
    &&\sum_{i=2}^{n}\tau_{k}[i]P_{k}^{\text{UL}}[i] \leq \frac{1}{\delta_{N}}\sum_{i=1}^{n-1}E_{k}^{L}[i](z_{k}[i] \mid \hat{z}_{k}[i]),\ \mbox{for} \ \ n\in\mathcal{\hat{N}} \ \mbox{and} \ k\in\mathcal{K}, \label{eq:P1approx-AvailEn}   \\
    && \mbox{(\ref{eq:P1-Speed}) - (\ref{eq:P1-MaxULPower}), (\ref{eq:P1eq-Aux})}. \nonumber
\eea
(P1.1A) can be solved by existing convex solvers, e.g., CVX \cite{CVX:17}. Since the feasible region of (P1.1A) is a subset of that of the original problem (P1.1), we can always obtain a lower bound solution for problem (P1.1) from its approximation (P1.1A).

As a result, a solution for (P1.1) can be calculated by iteratively solving (P1.1A) based on the CCCP. At the $i$-th iteration of the CCCP algorithm, we compute the solution $z_{k}^{(i)}[n]$ and $P_{k}^{\text{UL}(i)}[n]$ of (P1.1A) by setting $\hat{z}_{k}[n]=z_{k}^{(i-1)}[n]$, where $z_{k}^{(i)}[n]$ and $P_{k}^{\text{UL}(i)}[n]$ are the solution determined at the $i$-th iteration. In this algorithm, we set ${z}_{k}^{(0)}[n]$ to ${z}_{k}^{(0)}[n]=\|\mathbf{p}[n]-\mathbf{u}_{k}\|^2$ for all $n\in\mathcal{N}$ and $k\in\mathcal{K}$. It has been proved that this CCCP method converges to at least a local optimal point \cite{LAn:05}.
Note that for solving (P1.1) with the CCCP, we need to carefully initialize $\{\mathbf{p}[n]\}$. This will be clearly explained in Section \ref{sec:3-3_InitAP}.

\subsection{Time Resource Allocation} \label{sec:3-2_subprob2}
Now, we identify a solution for the time resource allocation $\{\tau_{k}[n]\}$ for given $\{\mathbf{p}[n]\}$ and $\{P_{k}^{\text{UL}}[n]\}$. The problem is written as
\bea
    \lefteqn{\mbox{(P1.2) }\max_{R_{\text{min}},\{\tau_{k}[n]\}}R_{\text{min}}} \label{eq:P1-2} \nonumber \\
    \lefteqn{s.t. }
    &&\frac{1}{N}\sum_{n=2}^{N}A_{k}[n]\tau_{k}[n] \geq R_{\text{min}},\ \mbox{for} \ \  k\in\mathcal{K}, \label{eq:P1-2-Throughput}\\ 
    &&\sum_{i=2}^{n}P_{k}^{\text{UL}}[i]\tau_{k}[i]
                        \leq \sum_{i=1}^{n-1}B_{k}[i]\tau_{0}[i],\ \ \mbox{for} \ \ n\in\mathcal{\hat{N}} \ \mbox{and} \ k\in\mathcal{K}, \label{eq:P1-2-AvailEn}  \\ 
    && \mbox{(\ref{eq:P1-tau1}) - (\ref{eq:P1-tau2})}, \nonumber
\eea
where $A_{k}[n]\triangleq\log_{2}\big(1+\frac{g_{0}\eta}{\sigma^2}\frac{P_{k}^{\text{UL}}[n]}{\|\mathbf{p}[n]-\mathbf{u}_{k}\|^2 + H^2}\big)$ and $B_{k}[n]\triangleq\frac{g_{0}\zeta_{k}P^{\text{DL}}}{\|\mathbf{p}[n]-\mathbf{u}_{k}\|^2 + H^2}$.
It can be shown that (P1.2) is a convex LP, which can be optimally solved by the standard LP optimization tools.

\renewcommand{\arraystretch}{1.1}
\begin{center}
\begin{tabular}{l}
\\
\hthickline
$\mathbf{Algorithm}$ $\mathbf{1}$: Proposed Algorithm for (P1)\\
\hthickline
Initialize\makebox[10pt]{   }$\tau_{k}^{(q)}[n]$ and $\mathbf{p}^{(q)}[n]$, $\forall n$ and $\forall k$, and set $q=0$. \\
Repeat\\
\makebox[10pt]{   } Set $q \leftarrow q+1$, $i \leftarrow 0$ and $z_{k}^{(q,i)}[n]=\|\mathbf{p}^{(q-1)}[n]-\mathbf{u}_{k}\|^2$, $\forall n$ and $\forall k$. \\
\makebox[10pt]{   } Repeat\\
\makebox[25pt]{   } Set $\hat{z}_{k}[n] = z_{k}^{(q,i)}[n]$, $\forall n$ and $\forall k$.\\
\makebox[25pt]{   } Solve (P1.1A) 
for given $\{\tau_{k}^{(q-1)}[n]\}$ by using the CVX. \\
\makebox[25pt]{   } Update $i \leftarrow i+1$. \\
\makebox[10pt]{   } Until convergence. \\
\makebox[10pt]{   } Update $\mathbf{p}^{(q)}[n]=\mathbf{p}^{(q,i)}[n]$ and $P_{k}^{\text{UL}(q)}[n]=P_{k}^{\text{UL}(q,i)}[n]$, $\forall n$ and $\forall k$.\\
\makebox[10pt]{   } Compute $R_{\text{min}}^{(q)}$ and $\{\tau_{k}^{(q)}[n]\}$ from (P1.2) for given $\{\mathbf{p}^{(q)}[n]\}$ and $\{P_{k}^{\text{UL}(q)}[n]\}$.\\
Until $R_{\text{min}}^{(q)}$ converges.\\
\hthickline \\
\end{tabular}
\end{center}

As a result, a solution of (P1) can be obtained by employing the alternating optimization framework and the overall process is given in Algorithm 1. In this algorithm, (P1.1) and (P1.2) are iteratively solved by fixing $\{\tau_{k}[n]\}$ and $\{\mathbf{p}[n]$, $P_{k}^{\text{UL}}[n]\}$, respectively. To be specific, at the $q$-th iteration, Algorithm 1 first successively solves (P1.1A) for given $\{\tau_{k}^{(q-1)}[n]\}$ based on the CCCP until the objective value converges. Note that we denote the solution obtained at the $i$-th iteration of the CCCP method as $\{\mathbf{p}^{(q,i)}[n], P_{k}^{\text{UL}(q,i)}[n],z_{k}^{(q,i)}[n]\}$. Then, a solution of (P1.2) is computed for given $\{\mathbf{p}^{(q)}[n],P_{k}^{\text{UL}(q)}[n]\}$, and this procedure is repeated until convergence.


Now, we verify the convergence of Algorithm 1. Let us define $\tilde{R}_{\text{min}}^{(q)}$ and $R_{\text{min}}^{(q)}$ as the objective value from the CCCP for (P1.1) and the optimal value of (P1.2) at the $q$-th iteration, respectively. Then, it is obvious that $R_{\text{min}}^{(q)} \leq \tilde{R}_{\text{min}}^{(q+1)}$ since the CCCP algorithm monotonically increases the objective value of (P1.1A) with respect to the iteration index $i$. Also, due to the fact that $R_{\text{min}}^{(q+1)}$ is the global optimal value of (P1.2) for given $\{\mathbf{p}^{(q+1)}[n]\}$ and $\{P_{k}^{\text{UL}(q+1)}[n]\}$, it follows $\tilde{R}_{\text{min}}^{(q+1)} \leq R_{\text{min}}^{(q+1)}$.

As a result, we have
\bea
    R_{\text{min}}^{(q)} \leq \tilde{R}_{\text{min}}^{(q+1)} \leq R_{\text{min}}^{(q+1)}, \nonumber\label{eq:convergence}
\eea
which implies that $R_{\text{min}}^{(q)}$ is non-decreasing with respect to the iteration index $q$. Because the minimum throughput $R_{\text{min}}$ is upper-bounded by a certain value, Algorithm 1 is guaranteed to converge. It is worth noting that the solutions at each iteration of Algorithm 1 are given by the local optimum and the global optimum for (P1.1) and (P1.2), respectively. For this reason, Algorithm 1 always yields at least a local optimal point for (P1).

\subsection{Trajectory Initialization} \label{sec:3-3_InitAP}
In this subsection, we present a simple initialization method for Algorithm 1. Although the time resource allocation $\{\tau_{k}[n]\}$ satisfying (\ref{eq:P1-tau1}) and (\ref{eq:P1-tau2}) can be initialized without problems, it is not easy to determine the feasible initial trajectory $\{\mathbf{p}[n]\}$ due to the complicated constraints in (\ref{eq:P1-Speed}) and (\ref{eq:P1-InitFinalLoc}). Thus, we apply the circular path scheme in \cite{QWu:17} to our scenario whose center $\mathbf{c}=[x_{c}\mbox{ }y_{c}]^{T}$ and radius $r$ on xy-plane are respectively set to
\bea
    \mathbf{c} &\triangleq& \frac{1}{K}\sum_{k=1}^{K}\mathbf{u}_{k}, \nonumber  \\
    r &\triangleq& \max(r^{\text{md}},r^{\text{max}}), \nonumber
\eea
where $\mathbf{c}$ represents the centroid of the GTs, and $r^{\text{md}}\triangleq\frac{1}{K}\sum_{k=1}^{K}\|\mathbf{c}-\mathbf{u}_{k}\|$ and $r^{\text{max}}\triangleq\frac{v_{\text{max}}T}{2\pi}$ indicate the mean distance between $\mathbf{c}$ and the GTs and the maximum allowable radius of the path with given speed constraint $v_{\text{max}}$, respectively. Thereby, the initial UAV trajectory becomes
\bea
  \mathbf{p}[n] = [x_{c}+r\cos{\frac{2\pi n}{N}} \ \ y_{c}+r\sin{\frac{2\pi n}{N}}]^{T}, \ \mbox{for} \ n\in\mathcal{N}. \label{initPath}
\eea
Note that for Algorithm 1, the time resource allocation $\{\tau_{k}[n]\}$ should also be initialized. The details will be discussed in Section \ref{sec:5-0simulation}.

\section{Proposed Solution for Separated UAV WPCN} \label{sec:4-0PropSol_SS}
In this section, we present an efficient algorithm for (P2) based on the alternating optimization. Similar to the integrated UAV WPCN, we first finds a solution for the trajectories $\{\mathbf{p}_{\scriptscriptstyle{\text{I}}}[n]\}$ and $\{\mathbf{p}_{\scriptscriptstyle{\text{E}}}
[n]\}$ and the uplink power $\{P_{k}^{\text{UL}}[n]\}$ for given $\{\tau_{k}[n]\}$, and then computes $\{\tau_{k}[n]\}$ for fixed $\{\mathbf{p}_{\scriptscriptstyle{\text{I}}}[n]\}$, $\{\mathbf{p}_{\scriptscriptstyle{\text{E}}}[n]\}$, and $\{P_{k}^{\text{UL}}[n]\}$. The details are described in the following subsections.

\subsection{Joint Trajectories and Uplink Power Optimization} \label{sec:4-1_subprob1}
In this subsection, we optimize $\{\mathbf{p}_{\scriptscriptstyle{\text{I}}}[n]\}$, $\{\mathbf{p}_{\scriptscriptstyle{\text{E}}}[n]\}$, and $\{P_{k}^{\text{UL}}[n]\}$ with given $\{\tau_{k}[n]\}$. In this case, (P2) can be simplified as
\bea
    \lefteqn{\max_{ R_{\text{min}}, \{P_{k}^{\text{UL}}[n]\}, \{\mathbf{p}_{\scriptscriptstyle{\text{I}}}[n]\}, \{\mathbf{p}_{\scriptscriptstyle{\text{E}}}[n]\} }R_{\text{min}} \quad \quad \quad \quad \quad \quad \quad \quad \quad } \label{P2_sub1} \\
    &s.t. \quad \quad \mbox{(\ref{eq:P2-Throughput}) - (\ref{eq:P2-MaxULPower})}. \nonumber
\eea
To solve the non-convex problem (\ref{P2_sub1}), similar to (P1.1), we introduce new auxiliary variables $\{z_{k}^{\text{I}}[n]\}$ and $\{z_{k}^{\text{E}}[n]\}$ such that $\|\mathbf{p}_{\scriptscriptstyle{\text{I}}}[n]-\mathbf{u}_{k}\|^2 \leq z_{k}^{\text{I}}[n]$ and $\|\mathbf{p}_{\scriptscriptstyle{\text{E}}}
[n]-\mathbf{u}_{k}\|^2 \leq z_{k}^{\text{E}}[n]$ for $k\in\mathcal{K}$ and $n\in\mathcal{N}$.

Then, problem (\ref{P2_sub1}) can be reformulated as
\bea
    \lefteqn{\mbox{(P2.1) }\max_{\substack{R_{\text{min}}, \{P_{k}^{\text{UL}}[n]\}, \{\mathbf{p}_{\scriptscriptstyle{\text{I}}}[n]\}, \\ \{\mathbf{p}_{\scriptscriptstyle{\text{E}}}[n]\},\{z_{k}^{\text{I}}[n]\},\{z_{k}^{\text{E}}[n]\}}}R_{\text{min}}} \nonumber \\
    \lefteqn{s.t. }
    &&\frac{1}{N}\sum_{n=2}^{N}\tau_{k}[n]\log_{2}\bigg(1+\frac{g_{0}\eta}{\sigma^2}\frac{P_{k}^{\text{UL}}[n]}{z_{k}^{\text{I}}[n] + H_{\text{I}}^2}\bigg)\geq R_{\text{min}},\ \mbox{for} \ \ k\in\mathcal{K}, ~~~~~~~~~~~~~~~~~~~~\label{eq:P2-1E-Throughput}\\
    &&\sum_{i=2}^{n}\tau_{k}[i]P_{k}^{\text{UL}}[i]
                        \leq \sum_{i=1}^{n-1}\tau_{0}[i]\frac{g_{0}\zeta_{k}P^{\text{DL}}}{z_{k}^{\text{E}}[i] + H_{\text{E}}^2},\ \mbox{for} \ \ n\in\mathcal{\hat{N}} \ \mbox{and} \ k\in\mathcal{K},\label{eq:P2-1E-AvailEn}\\
    &&\|\mathbf{p}_{\scriptscriptstyle{\text{I}}}[n]-\mathbf{u}_{k}\|^2 \leq z_{k}^{\text{I}}[n],\ \mbox{for} \ \ n\in\mathcal{N} \ \mbox{and} \ k\in\mathcal{K}, \label{eq:P2-1E-AuxBS} \\
    &&\|\mathbf{p}_{\scriptscriptstyle{\text{E}}}[n]-\mathbf{u}_{k}\|^2 \leq z_{k}^{\text{E}}[n],\ \mbox{for} \ \ n\in\mathcal{N} \ \mbox{and} \ k\in\mathcal{K}, \label{eq:P2-1E-AuxPB} \\
    &&\mbox{(\ref{eq:P2-SpeedIAP}) - (\ref{eq:P2-MaxULPower})}. \nonumber
\eea
The equivalence between problem (\ref{P2_sub1}) and (P2.1) can be easily verified by a similar approach in Lemma 1 since we have $\|\mathbf{p}_{\scriptscriptstyle{\text{I}}}[n]-\mathbf{u}_{k}\|^2 = z_{k}^{\text{I}}[n]$ and $\|\mathbf{p}_{\scriptscriptstyle{\text{E}}}[n]-\mathbf{u}_{k}\|^2 = z_{k}^{\text{E}}[n]$ at the optimal point of (P2.1).

As in (P1.1), we can check that (P2.1) with the non-convex constraints (\ref{eq:P2-1E-Throughput}) and (\ref{eq:P2-1E-AvailEn}) is the difference of convex problem which can be handled by the CCCP method \cite{LAn:05}.
Thus, at each iteration of the CCCP algorithm, we address the following approximated convex problem as
\bea
    \lefteqn{\mbox{(P2.1A) }\max_{\substack{R_{\text{min}}, \{P_{k}^{\text{UL}}[n]\}, \{\mathbf{p}_{\scriptscriptstyle{\text{I}}}[n]\}, \\ \{\mathbf{p}_{\scriptscriptstyle{\text{E}}}[n]\},\{z_{k}^{\text{I}}[n]\},\{z_{k}^{\text{E}}[n]\}}}R_{\text{min}}} \nonumber \\
    \lefteqn{s.t. }
    &&\frac{1}{N}\sum_{n=2}^{N}\tau_{k}[n]R_{k}^{L}[n](z_{k}^{\text{I}}[n],P_{k}^{\text{UL}}[n]\mid \hat{z}_{k}^{\text{I}}[n]) \geq R_{\text{min}},\ \mbox{for} \ k\in\mathcal{K}, \label{eq:P2approx-Throughput}\\
    &&\sum_{i=2}^{n}\tau_{k}[i]P_{k}^{\text{UL}}[i] \leq \frac{1}{\delta_{N}}\sum_{i=1}^{n-1}E_{k}^{L}[i](z_{k}^{\text{E}}[i] \mid \hat{z}_{k}^{\text{E}}[i]),\ \mbox{for} \ \ n\in\mathcal{\hat{N}} \ \mbox{and} \ k\in\mathcal{K}, \label{eq:P2approx-AvailEn} \\
    &&\mbox{(\ref{eq:P2-SpeedIAP}) - (\ref{eq:P2-MaxULPower})},\mbox{(\ref{eq:P2-1E-AuxBS}) - (\ref{eq:P2-1E-AuxPB})}, \nonumber
\eea
where the approximations in (\ref{eq:P2approx-Throughput}) and (\ref{eq:P2approx-AvailEn}) are obtained from (\ref{Rlower}) and (\ref{Elower}), respectively.
Therefore, we can compute a local optimal solution for (P2.1) by iteratively solving (P2.1A) with $\hat{z}_{k}^{\text{I}}[n] = z_{k}^{I(i)}[n]$ and $\hat{z}_{k}^{\text{E}}[n] = z_{k}^{E(i)}[n]$ for $n\in\mathcal{N}$ and $k\in\mathcal{K}$, whose convergence has been shown in \cite{LAn:05}. The overall procedure is similar to that for (P1.1) and thus omitted here for brevity.

\subsection{Time Resource Allocation} \label{sec:4-2_subprob2}
With given $\{\mathbf{p}_{\scriptscriptstyle{\text{I}}}[n]\}$, $\{\mathbf{p}_{\scriptscriptstyle{\text{E}}}[n]\}$, and $\{P_{k}^{\text{UL}}[n]\}$, we now determine the time resource allocation solution $\{\tau_{k}[n]\}$. The problem is written as
\bea
    \lefteqn{\mbox{(P2.2) }\max_{R_{\text{min}},\{\tau_{k}[n]\}}R_{\text{min}}} \nonumber \label{eq:P2-2}\\
    \lefteqn{s.t. }
    &&\frac{1}{N}\sum_{n=2}^{N}A_{k}^{\text{I}}[n]\tau_{k}[n] \geq R_{\text{min}},\ \mbox{for} \ \  k\in\mathcal{K}, ~~~~~~~~~~~~~~~~~~~~\label{eq:P2-2-Throughput}\\ 
    &&\sum_{i=2}^{n}P_{k}^{\text{UL}}[i]\tau_{k}[i]
                        \leq \sum_{i=1}^{n-1}B_{k}^{\text{E}}[i]\tau_{0}[i],\ \ \mbox{for} \ \ n\in\mathcal{\hat{N}} \ \mbox{and} \ k\in\mathcal{K},\label{eq:P2-2-AvailEn}  \\ 
    && \mbox{(\ref{eq:P1-tau1}) - (\ref{eq:P1-tau2})}, \nonumber
\eea
where $A_{k}^{\text{I}}[n] \triangleq \log_{2}\big(1+\frac{g_{0}\eta}{\sigma^2}\frac{P_{k}^{\text{UL}}[n]}{\|\mathbf{p}_{\scriptscriptstyle{\text{I}}}[n]-\mathbf{u}_{k}\|^2 + H^2}\big)$ and $B_{k}^{\text{E}}[n] \triangleq \frac{g_{0}\zeta_{k}P^{\text{DL}}}{\|\mathbf{p}_{\scriptscriptstyle{\text{E}}}
[n]-\mathbf{u}_{k}\|^2 + H^2}$. One can check that (P2.2) is a convex LP which can be optimally solved by the standard LP solver.

Finally, we can obtain a local optimal solution of (P2) by applying the alternating optimization where we first identify a solution of (P2.1) via the CCCP algorithm, and then update $\{\tau_{k}[n]\}$ from (P2.2). This procedure is repeated until the minimum throughput converges. The convergence and the local optimality of such a process can be directly verified by following the convergence proof of Algorithm 1, and thus are omitted. Also, for initializing the trajectories of two UAVs $\{\mathbf{p}_{\scriptscriptstyle{\text{I}}}[n]\}$ and $\{\mathbf{p}_{\scriptscriptstyle{\text{E}}}[n]\}$, we can adopt the circular trajectory scheme presented in Section \ref{sec:3-3_InitAP}.

\section{Simulation Results} \label{sec:5-0simulation}
\begin{figure}
    \centering
    \includegraphics[width=\FigureWidth]{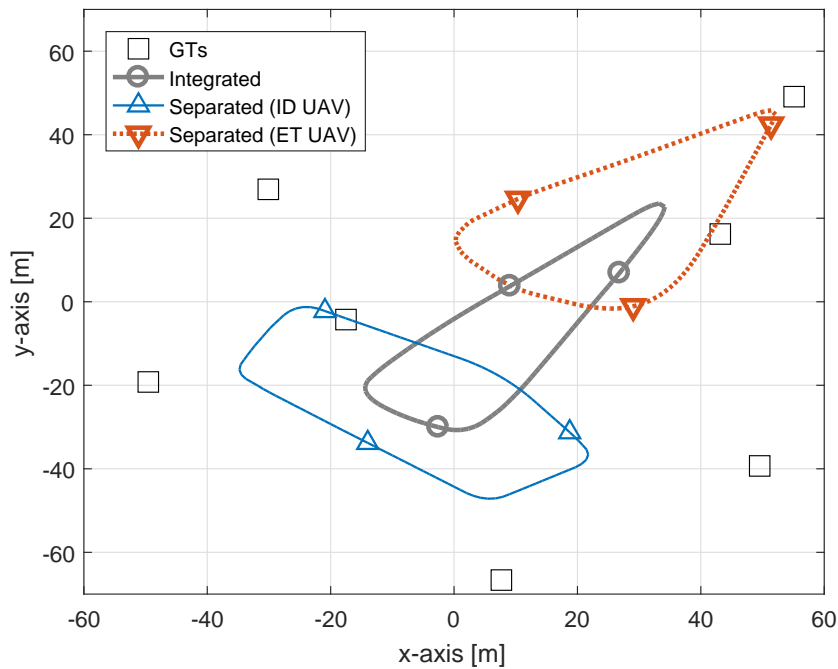}
    \caption{Trajectories of UAVs optimized by the proposed algorithms for $T=30$ sec.}
    \label{fig:traj_nonfixed_T30}
\end{figure}

In this section, we evaluate the performance of the proposed algorithms by numerical results. Unless stated otherwise, the maximum uplink power constraint at the GTs and the downlink transmission power at the UAVs are equal to $P_{\text{max}}^{\text{UL}} = -10$ dBm and $P^{\text{DL}} = 30$ dBm, respectively. Also, $\zeta$ and $\eta$ are fixed as $\zeta=0.6$ and $\eta=0.9$, respectively. We set the reference channel gain $g_{0}$ to $g_{0}=-30$ dB and the noise variance is given by $\sigma^{2}=-90$ dBm. The speed of the UAVs is limited to $v_{\text{max}}=5$ m/s and the altitude of the UAVs is fixed to $H=H_{\text{I}}=H_{\text{E}}=8$ m.
For simulations, we adopt a system with seven GTs ($K=7$), whose locations are marked by squares in Figure \ref{fig:traj_nonfixed_T30}. The initial time resource allocation $\{\tau_{k}[n]\}$ for the proposed algorithms is determined as $\tau_{k}[n]=\frac{1}{K+1}$, $\forall n$ and $\forall k$.

\begin{figure}
  \centering
  \includegraphics[width=\FigureWidth]{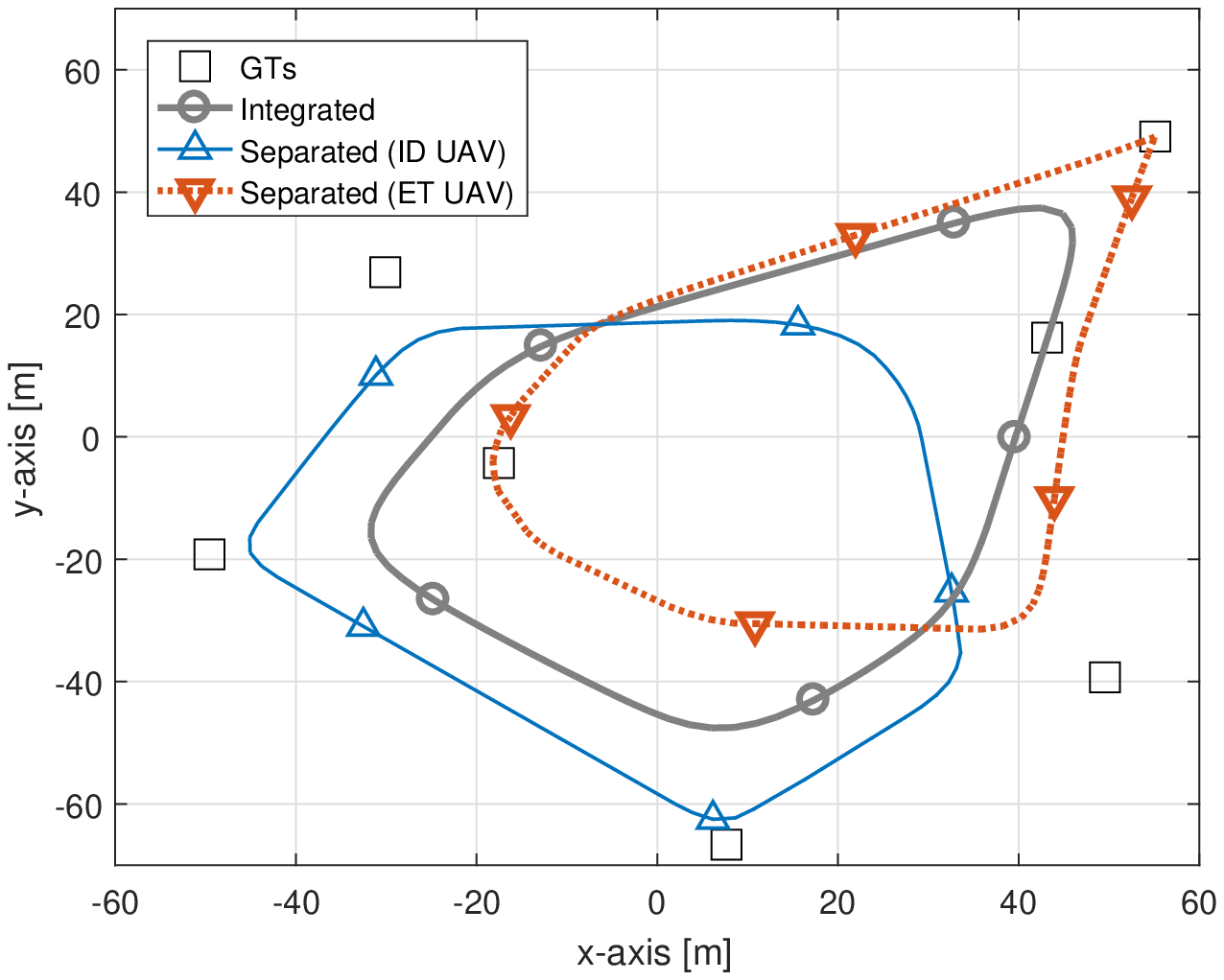}
  \caption{Trajectories of UAVs optimized by the proposed algorithms for $T=50$ sec.}
  \label{fig:traj_nonfixed_T50}
\end{figure}

\begin{figure}
\centering
    \includegraphics[width=\FigureWidth]{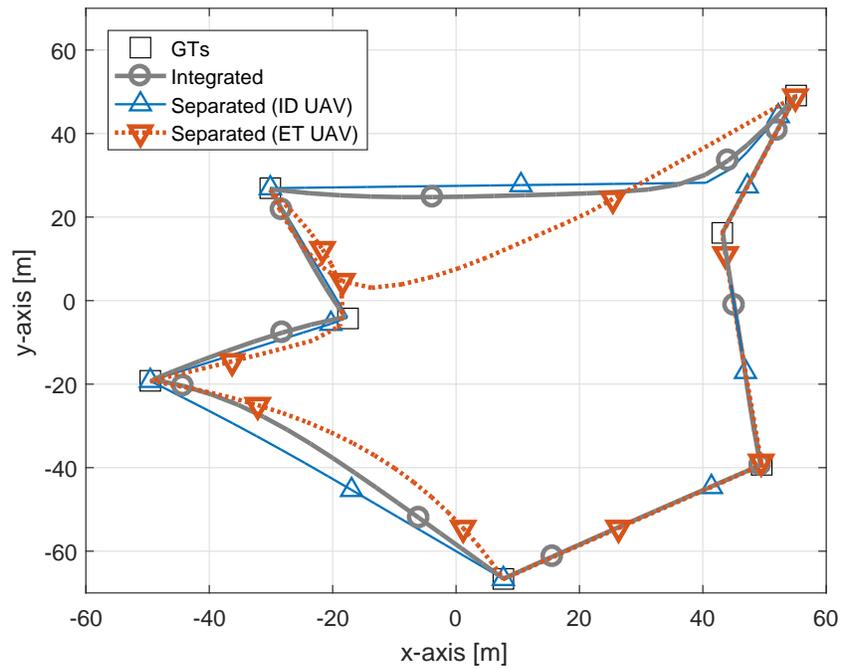}
    \caption{Trajectories of UAVs optimized by the proposed algorithms for $T=100$ sec.}
    \label{fig:traj_nonfixed_T100}
\end{figure}

Figures \ref{fig:traj_nonfixed_T30}-\ref{fig:traj_nonfixed_T100} illustrate the optimized trajectories of the UAVs in the integrated UAV and the separated UAV WPCNs for $T=30,50$, and $100$ sec, respectively. The triangular and circular markers represent the positions of UAVs sampled at every $10$ sec.
First, in Figure \ref{fig:traj_nonfixed_T30}, we can see that the UAV in the integrated UAV WPCN tries to cover all GTs by traveling a path whose center is close to the centroid of the GTs. In contrast, in the separated system, two UAVs mainly cover two different areas so that the ET UAV flies over the upper-right side of the area, while the ID UAV gets around the lower-left side. This is because when $T=30$ sec, the time period is not enough for two UAVs to visit all the GTs. Therefore, by sectorizing the area, the separated UAV WPCN can transfer energy and receive information more efficiently compared to the integrated WPCN. This tendency also can be observed in Figure \ref{fig:traj_nonfixed_T50}.
Note that the minimum throughput performance of the separated UAV WPCN is 51\% and 31\% larger than that of the integrated UAV WPCN for $T=30$ and $50$ sec, respectively.

In Figure \ref{fig:traj_nonfixed_T100}, the optimized trajectory for the integrated UAV WPCN converges to a path consisting of line segments which connect the locations of the GTs for large $T$ as in \cite{QWu:17}. Nevertheless, the separated UAV WPCN still exhibits non-trivial trajectories due to the decoupled WET and WIT operations. When $T=100$ sec, the minimum throughput performance of the separated UAV WPCN is about 15\% larger than that of the integrated UAV WPCN.

\begin{figure}
\begin{center}
\includegraphics[width=\FigureWidth]{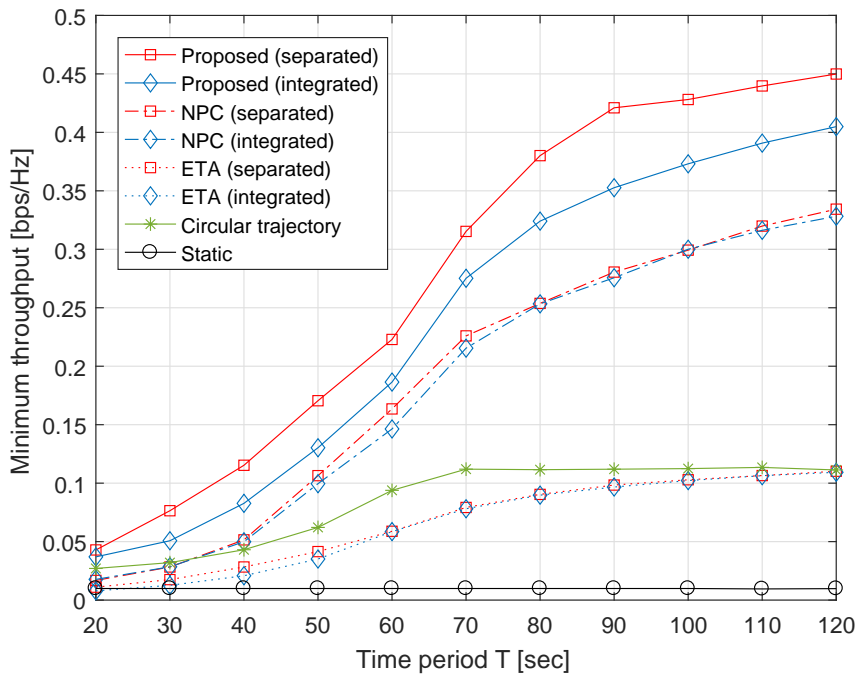}
\end{center}
\vspace{0mm}
\caption{Minimum throughput with respect to time interval $T$ for various systems.}
\label{fig:Tsweep}
\end{figure}

Figure \ref{fig:Tsweep} illustrates the minimum throughput performance of the proposed algorithms with respect to the time interval $T$. For comparison, we also plot the performance of the following baseline schemes.
\begin{itemize}
  \item Static AP: The conventional WPCN \cite{HJu:14} with a static H-AP is adopted where the H-AP is fixed at the centroid of GTs with the altitude of $H=8$ m. The time resource allocation solution is optimized from the algorithm presented in \cite{HJu:14}.
  \item Circular trajectory: The UAVs follow the circular path presented in Section \ref{sec:3-3_InitAP}. Then, the uplink power and the time resource allocation are optimized from Algorithm 1 with fixed $\{\mathbf{p}[n]\}$ in (\ref{initPath}). Note that with the circular trajectory, both the integrated UAV and the separated UAV WPCNs achieve the same minimum rate performance.
  \item Equal time allocation (ETA): With equal time resource allocation $\tau_{k}[n]=\frac{1}{K+1}$, $\forall n$ and $\forall k$, the uplink power and the trajectory of the UAV are obtained from (P1.1) by applying the CCCP algorithm explained in Section \ref{sec:3-1_subprob1}.
  \item Naive power control (NPC): With the trajectory and the time resource allocation optimizations based on the proposed algorithms, each GT uses all of the energy harvested at the previous time slot for WIT. If the stored energy at the GT exceeds the uplink power constraint, i.e., $E_k[n-1]>\delta_{N}\tau_{k}[n]P^{\text{UL}}_{\text{max}}$, the GT transmits the information signal with $P^{\text{UL}}_{\text{max}}$ at the $n$-th time slot.
\end{itemize}
From Figure \ref{fig:Tsweep}, we can check that even when the trajectory is simply set to the circular path, the minimum throughput can be improved compared to the conventional static WPCN. Also, the ETA case shows a performance enhancement by optimizing the trajectory of the UAV without the time resource allocation.
These infer that the mobility of the UAV well compensates the doubly near-far problem of the static WPCN.
Although the NPC scheme naively controls the uplink power, the minimum throughput is further improved in comparison with other baseline schemes by jointly optimizing the UAV trajectories and the time resource allocation.
The minimum throughput performance of the proposed algorithms increases as the time period $T$ grows, and the performance increment becomes smaller for a large $T$.
In addition, the performance gap between the proposed schemes and the conventional methods grows with $T$.
Note that this indicates that the optimization of the UAV trajectories and the time resource allocation can bring a huge gain on system performance, and thus these are critical design factors.
Moreover, the separated UAV WPCN always performs better than the integrated UAV WPCN in the proposed scheme, while the baseline schemes do not exhibit such advantages. This can be attributed to a fact that in the NPC and the ETA schemes, both the ID and the ET UAV trajectories in the separated system converge to the same trajectory, due to the limited energy causality and time resource allocation.
Therefore, we can conclude that jointly optimization of trajectories, uplink power control and time resource allocation is important for the UAV WPCNs.

\begin{figure}
\begin{center}
\includegraphics[width=\FigureWidth]{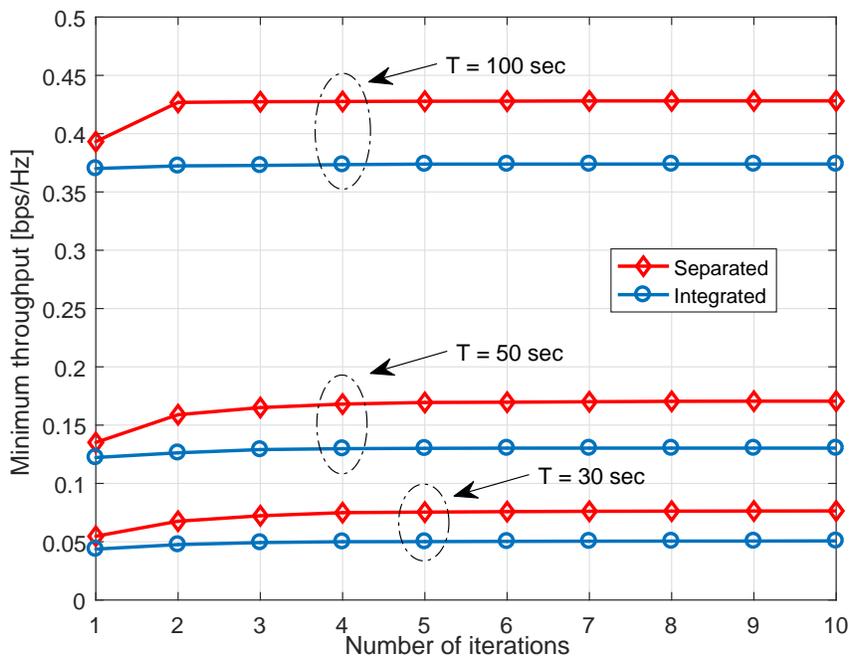}
\end{center}
\vspace{0mm}
\caption{Convergence of the proposed algorithms.}
\label{fig:convergence}
\end{figure}

In Figure \ref{fig:convergence}, we show the convergence of the proposed algorithms for the UAV WPCNs. We observe that for all time periods, the proposed algorithms for the integrated UAV WPCN seem to converge within one iteration, while that for the separated UAV WPCN requires about three iterations for convergence.
%

\section{Conclusion} \label{sec:6-0conclusion}
This paper has investigated the UAV WPCN where mobile UAVs support the WET and the WIT of multiple GTs. For both the integrated UAV and the separated UAV WPCNs, we have jointly optimized the trajectories of the UAVs, the uplink power at the GTs, and the time resource allocation strategies in order to maximize the minimum throughput among the GTs. To solve these non-convex problems, we have adopted the alternating optimization framework and the CCCP algorithm. As a result, a locally optimal solution of the original non-convex problems has been efficiently computed by the proposed algorithms whose convergence has been mathematically proved. From the simulation results, we have demonstrated the efficiency of the proposed algorithms over the conventional schemes.

\bibliographystyle{ieeetr}
\input{bibliography.filelist}

\end{document}